# On the Properties of Random Multiplicative Measures with the Multipliers Exponentially Distributed

Wei-Xing Zhou[*], and Zun-Hong Yu

*East China University of Science and Technology, P. O. Box 272, Shanghai 200237, P.R.China*

*Corresponding author. Fax: 86-21-64250192. Tel: 86-21-64253933. E-mail: wxzhou@ecust.edu.cn

**Abstract**

Under the formalism of annealed averaging of the partition function, a type of random multifractal measures with their multipliers satisfying exponentially distributed is investigated in detail. Branching emerges in the curve of generalized dimensions, and negative values of generalized dimensions arise. Three equivalent methods of classification of the random multifractal measures are proposed, which is based on: (i) the discrepancy between the curves of generalized dimensions, (ii) the solution properties of equation $\tau(q) = 0$, and (iii) the relative position of the curve $f(\alpha)$ and the diagonal $f(\alpha) = \alpha$ in the first quadrant. These three classes correspond to $\mu([0,1]) = \infty$, $\mu([0,1]) = 1$ and $\mu([0,1]) = 0$, respectively. Phase diagram is introduced to illustrate the diverse performance of the random measures that is multiplicatively generated.

*PACS:* 05.45.Df, 47.53.+n, 02.50.Ey

*Keywords:* Random multifractals; Exponential distribution; Classification; Branching; Phase diagram

In the formalism of either the restricted theory of multifractal [1-5] or the general theory of multifractal [6-10], multifractal measures are decomposed into interwoven fractal sets each of which is characterized by its singularity. Following Mandelbrot [6], the earlier and more general meaning of the term "multifractal" comes from the notation of "multiplicative cascade that generates nonrandom or random measures", and describes "measures that are multiplicatively generated" [11]. In the deterministic case, the singularity spectrum $f(\alpha)$ is always non-negative and varies in a finite range [12,13]. However, when one investigates the multifractal nature of random measures arising from experiments, say the diffusion-limited aggregation [14], diffusion [15] and the dissipation field of turbulence [16], negative dimensions are discovered, which means that $f(\alpha)$ can be negative for certain $\alpha$. The meaning of negative $f(\alpha)$ was discussed by Cates and Witten [15] for the first time. Later, with a simple but cogitative analytical example, Chhabra and Sreenivasan [17] pointed out that, the negative dimensions arise from either the intrinsic randomness or a random view of a deterministic process. Relative studies refer to Ref. [18-21].

A random multiplicative cascade process will generate a random multifractal measure on certain geometry support, which is a stochastic object. Two averaging is possible when investigating the scaling properties of such stochastic objects. One can choose to define either an annealed scaling exponent or a quenched one, respectively. Halsey has expected the quenched averaging to yield a more physically meaningful result in DLA [22,23]. In the formalism of quenched averaging, the scaling properties obtained are similar to the results in the deterministic case, where, say, $f(\alpha)$ is connaturally non-negative. Henceforth, the quenched averaging cuts off the intrinsic or practically deduced randomness for many processes, say the fully developed turbulence, where the annealed averaging shows its advantage in characterizing the lacunarity of certain occasionally emerging measures [17,24].

To calculate $f(\alpha)$ accompanied with the negative part, a universal procedure follows the scaling of histograms of the rescaled probability $p(\alpha)$ [20,21,25]. A more efficient procedure named multiplier method that can extract the $f(\alpha)$ spectrum with exponential less work and is more accurate than the conventional box-counting method was proposed in [17]. These studies emphasize particularly the experimental facets and focus mainly on the conservative cascades. In this paper, we present an analytic example of random multifractal with its multipliers exponentially distributed, which exhibit many interesting properties as well as "anomalous" behaviors. Particularly, non-conservative cascades [19,21] play an important roll in this paper.

First, perform a random multiplicative cascade process. Divide uniformly the interval $[0,1]$ into $b$ pieces with the multipliers $M$ whose probability density $\Pr(M)$ is continuous. This procedure is continued *ad* infinity. It is clear that the multiplicative process must produce a multifractal measure. From Cramer's theorem of large deviations [26-28], the mass exponent can be defined by an annealed averaging of moments of the multipliers, namely

$$\tau(q) = -D_0 - \frac{\log\langle M^q \rangle}{\log b}. \tag{1}$$

The generalized dimensions are defined by

$$D_q = \lim_{q \to \tilde{q}} \frac{\tau(\tilde{q})}{\tilde{q}-1}, \tag{2}$$

where $D_0$ is the fractal dimension of the geometric support and consequently equal to 1 in the present case. The multifractal spectrum $f(\alpha)$ is linked with $\tau(q)$ by Legendre transform and inverse Legendre transform [24]

$$\alpha(q) = \tau'(q) = -\frac{\langle M^q \log M \rangle}{\langle M^q \rangle \log b}, \tag{3}$$

and

$$f(\alpha(q)) = q\alpha(q) - \tau(q) = \frac{\langle M^q \rangle \log \langle M^q \rangle - \langle M^q \log M^q \rangle}{\langle M^q \rangle \log b} + D_0. \tag{4}$$

Now, consider a class of random measures with the multipliers satisfying the exponential distribution

$$\Pr(M) = \frac{\log x}{x-1} x^M, \tag{5}$$

where $x > 0$ and $0 < M < 1$. Therefore,

$$\langle M^q \rangle = \frac{\log x}{x-1} \int_0^1 M^q x^M \, dM . \tag{6}$$

The definition domain is $q > -1$. To obtain the expression analytically, we expand $x^M$ with power series and obtain

$$\langle M^q \rangle = \frac{\log x}{x-1} \cdot \sum_{n=0}^{\infty} \frac{(\log x)^n}{n!(n+q+1)} . \tag{7}$$

If $q$ is a non-negative integer, we have

$$\langle M^q \rangle = \frac{q!}{(x-1)(\log x)^q} \cdot \left[ (x-1)(-1)^q + x \sum_{n=1}^{q} \frac{(-1)^{n-1}(\log x)^{q+1-n}}{(q+1-n)!} \right] . \tag{8}$$

Since the ratio $\frac{(n+q)\log x}{n(n+q+1)}$ between the adjacent terms in (7) tends to zero for sufficiently large $n$, we can compute $\langle M^q \rangle$ via the truncated form of (7). An alternative way to compute $\langle M^q \rangle$ is to perform numerical integration according to (6). However, this procedure will cost exponentially more work with $q$ approaching to $-1$ [29]. Note that, if $x=1$, $\frac{\log(x)}{x-1} = 1$ in Equations (5-8). Illustrated in Figure 1-3 are the typical diagrams of the generalized dimensions in the cases of $x=e$, $x=1$ and $x=1/e$ accompanied with $b=2$. Figures 4-6 are the corresponding diagrams of $\tau(q)$, $\alpha(q)$ and $f(\alpha)$.

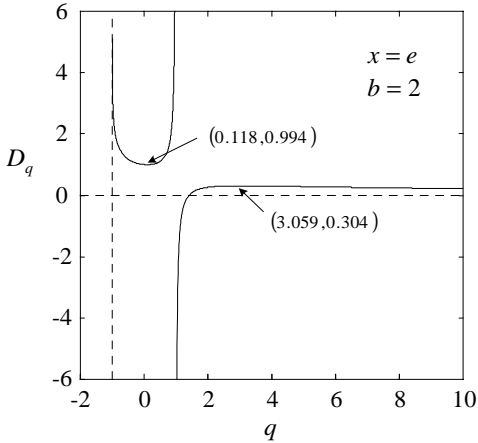 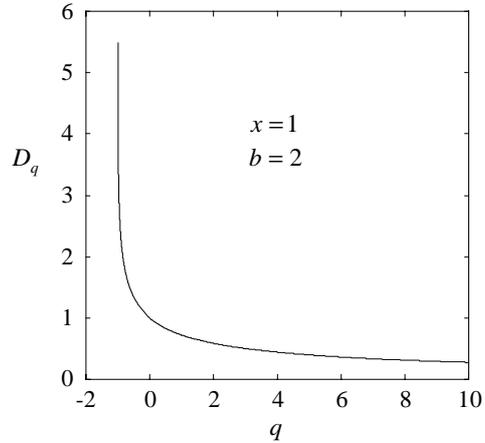

Fig. 1. Typical diagram of $D_q$ of Class I.    Fig. 2. Typical diagram of $D_q$ of Class II.

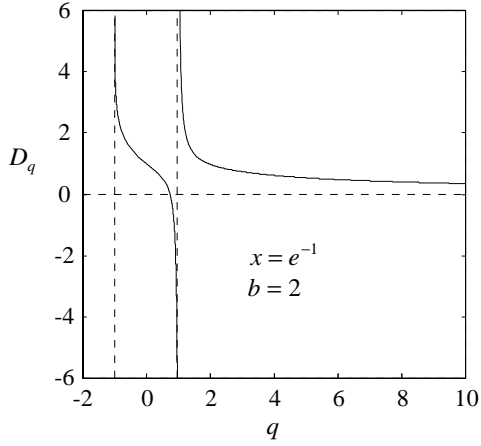

Fig. 3. Typical diagram of $D_q$ of Class III.

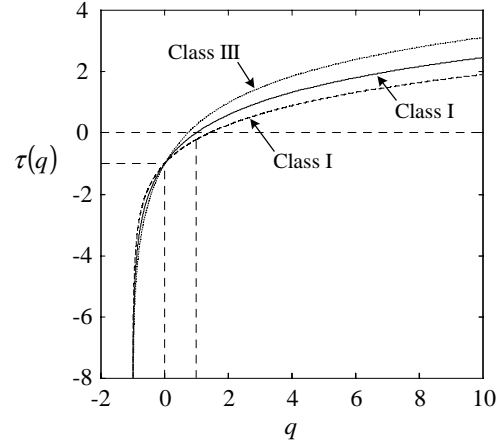

Fig. 4. Typical diagrams of $\tau(q)$.

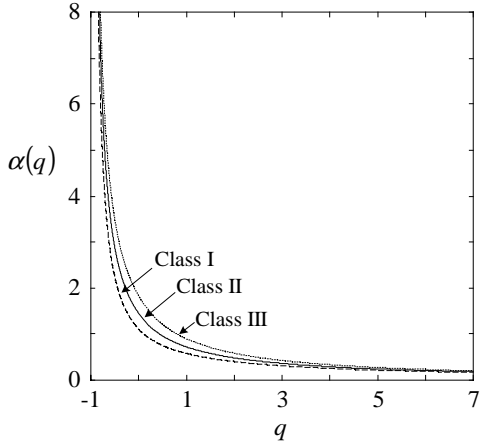

Fig. 5. Typical diagrams of $\alpha(q)$.

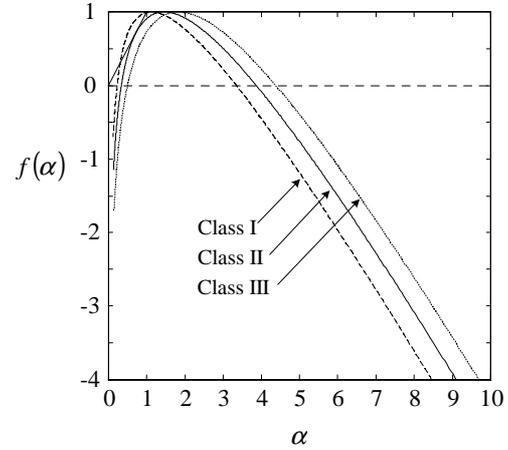

Fig. 6. Typical diagrams of $f(\alpha)$.

Obviously, there are three distinct classes according to the diagrams of generalized dimensions in Figures 1-3. *Class I*: there are two separated branches in the curve of generalized dimensions, each with an extreme point, as show in Figure 1. The two extreme values correspond to the solutions of the nonlinear equation $D'_q = 0$. *Class II*: there is a single continuous curve with no extreme value as shown in Figure 2. *Class III*: there are two separated monotonously decreasing branches with no extreme value as illustrated in Figure 3. Similar features are also discovered in the cases of power distribution and triangular distribution [30]. The cause of branching of generalized dimensions is due to two key points. One is about the solution nature of $D'_q = 0$, while the other is about the behavior of convergency of $D_q$ at point $q = 1$.

A more intuitive, but equivalent to the previous (as will be clarified hereafter), classification random multifractal measures is to investigate the relative position between the $f(\alpha)$ curve and the diagonal $f(\alpha) = \alpha$ of the first quadrant. *Class I*: Intersection. If the diagonal $f(\alpha) = \alpha$ intersects the $f(\alpha)$ curve, there must exist two intersecting points $(\alpha(q_1), f(q_1))$ and $(\alpha(q_2), f(q_2))$. Without loss of generality, we can regard that $q_1 < 1 < q_2$, since it is impossible that they intersect at $q = 1$ or the same side of $q = 1$. *Class II*: Tangency. This is the only case that one can use the so-called determining criterion [31] to judge whether the computed $f(\alpha)$ of random multifractal measures is valid or not. In this case we have $f(\alpha) \leq \alpha$ with equality at $q = 1$. *Class III*: Separation. In this case, the $f(\alpha)$ curve locates below the diagonal line $f(\alpha) = \alpha$, implying $f(\alpha) < \alpha$ for all $q$ and hence $\alpha$ as well. We also conjecture, in a more general sense, that the branching in the curve of generalized dimensions emerges again, and that no extreme points exist. Same situations are encountered when considering the relative position between the $f(q)$ curve and the $\alpha(q)$ curve.

Consider the solution $q_0$ of the nonlinear equation

$$\tau(q) = 0. \tag{9}$$

Since $\tau'(q) > 0$ according to (3) and $\tau(\pm\infty) = \pm\infty$ from (1), $q_0$ exists uniquely. From (4), we have $f(\alpha(q_0)) = q_0 \alpha(q_0)$, which is tangent to the $f(\alpha)$ curve at point $(\alpha(q_0), f(\alpha(q_0)))$ with the slope $q_0$ and passes through the Origin. Therefore, the situations of $q_0 > 1$, $q_0 = 1$ and $q_0 < 1$ are the counterparts of the cases of intersection, tangency and separation between the $f(\alpha)$ curve and the diagonal $f(\alpha) = \alpha$. From (2), we have

$$D_q' = (f - \alpha)/(q-1)^2. \tag{10}$$

If $q_0 > 1$, we find that $D_{q_1}' = D_{q_2}' = 0$, and hence $D_q' < 0$ for $-1 < q < q_1$ and $q > q_2$, while $D_q' > 0$ for $q_1 < q < 1$ and $1 < q < q_2$. If $q_0 < 1$, $D_q' < 0$ since $f < \alpha$ for any $q$ in the definition domain. If $q_0 = 1$, one can find that $D'_q < 0$ for any $q$, where $D'_1 = \alpha'(1)/2$ is used.

So far, the equivalency within these classification methods is clarified. Therefore, three equivalent rules are established to classify random multifractal measures, which come from the natures of $D_q$, $\tau(q)$ and $f(\alpha)$, respectively. It seems that one can't intuitively classify such measures via investigating the properties of $f(q)$ or $\alpha(q)$. The reason is because that, each of the $D_q$, $\tau(q)$ and $f(\alpha)$ can characterize fully the self-similarity of multifractal measures, and that they can transform from each one to others and are consequently equivalent to each other.

Consider the situation of $\tau(q) = 0$. The necessary and sufficient condition of $q_0 = 1$ is that

$$\langle M \rangle = b^{-1}. \tag{11}$$

This implies that the generated measure is of average conservation, also referred to as canonical measure [19,21,32,33], with $\mu([0,1]) = 1$ as the usual probability measures. From (8), we have

$$\langle M \rangle = \frac{x \log x - x + 1}{(x-1) \log x}. \tag{12}$$

Hence,

$$b = \frac{(x-1) \log x}{x \log x - x + 1}. \tag{13}$$

Since $\frac{db}{dx} < 0$, $b$ has a unique value for any fixed $x$, and vice versa. This also shows that $\langle M \rangle$ is increasingly monotone with $x$. We rewrite (11) in the form of its inverse function

$$x = h(b). \tag{14}$$

When $x = h(b)$, the random multifractal measures belong to Class II. When $x > h(b)$, the measures belong to Class I. Since $\langle M \rangle > b^{-1}$, we have $\mu([0,1]) = \infty$. When $x < h(b)$, the measures belong to Class III. Since $\langle M \rangle < b^{-1}$, we have $\mu([0,1]) = 0$. It is clear that measures in Class I and Class III are non-conservative. We can obtain the phase diagram of $(b, x)$ which is shown in Figure 7. The phase space is $\{(b,x): b > 1, x > 0\}$. The solid line II denotes measures in Class II, while regions I and III denotes measures in Class I and Class III, respectively. Equation

(14) represents the *critical curve*, which separates Class I and Class III. If only the integer bases $b$ are considered, these three regions degenerate to a group of parallel dotted radial lines (Class I), discrete solid dots (Class II), i.e., $(b, h(b))$ where $b = 2, 3, \cdots$, and parallel solid line segments (Class III). Formally, we suggest that non-integer base is also valid (like Mandelbrot's limit lognormal multifractal measures [34]) when investigating such random cascade process that there are different bases in different generations. Therefore, the non-integer base can be looked upon as an averaged base $\langle b \rangle$.

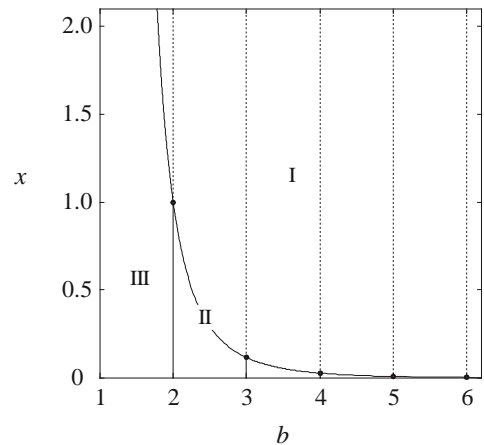

Fig. 7. Phase diagram arising from the random multiplicative measures with the multipliers $M$ exponentially distributed. Solid line II denotes measures in Class II, while regions I and III denotes measures in Class I and Class III, respectively. In the case of integer base $b$, only the parallel dotted radial lines (Class I), discrete solid dots (Class II) and parallel solid line segments (Class III) are validated.

We expect relevant application to fluid field [34,11] of non-conservative cascade, no matter it is deterministic or random. Falconer and O'Neil [35] had proposed a framework for vector-valued measures, which permits vectors analogues of the singular measures arising in multifractal theory. The formalism of vector-valued measures is same to the "ordinary" multifractals in the form. However, the multipliers, which they denoted as $t_i$ instead of $m_i$ as usual, are free from the constraints of $M<1$ and of conservation of measure. Hence, one can expect the vector-valued measures at infinite stage to be $\mu([0,1])=\infty$. Figure 5 in [35], which is an example of vector-valued measures, illustrates that the diagonal $f(\alpha)=\alpha$ intersects the $f(\alpha)$ curve. Such measures belong to Class I. The vector-valued measures can be used to describe the behavior of fluid, electric or magnetic fields and many other phenomena. In addition, such non-conservative multiplicative process is also expect to suit for modeling stratified resistance network [36] and size distribution of drop breakup in atomization process [37,38].

Negative generalized dimensions arise in Class I and III, which are caused by the definition of $D_q$ from $\tau(q)$. If $q_0>1$, then $\tau(q)<0$ for any $1<q<q_0$, and hence $D_q$ is negative. If $q_0=1$, $D_q$ is positive for all $q$ in the definition domain. If $q_0<1$, $\tau(q)>0$ for any $q_0<q<1$, and hence $D_q$ is negative. An intuitional view is demonstrated in Figures 1-3.

Generally, the tendencies of $D(q)$ and $\alpha(q)$ are similar to each other. Moreover, both the minimal and maximal values of $D(q)$ and $\alpha(q)$ exist and are identical respectively [11,12] in the restricted framework of multifractals, while about left-sided multifractal measures [6], the minimal $\alpha(q)$ exists and $\alpha \to +\infty$ when $q \to +\infty$. However, there are no boundaries for $\alpha(q)$ in its definition domain for the present case. As shown in Fig. 5, $\alpha \in (0,+\infty)$. Note that, $\alpha(q)$ cannot reach $0$ as its minimal value. The sufficient condition under which $\alpha_{\min}=0$ is that one can identify at least one point where $\alpha=0$. To meet $\alpha=0$, one should investigate the maximal measure $1$, which can never be attained in the present case since all multipliers are less than unity, though one can approach to it as near as possible. Therefore, in the case of discrete probability distribution of the multipliers, $\alpha_{\min}$ exists which corresponds to the region with maximal measure, but it is not the matter in the continuous case.

In summary, we have studied the multifractality of a random multiplicative measure with its multipliers $M$ exponentially distributed under the multifractal formalism of annealed averaging of the partition function. If the probability density of the random variable $M$ is continuous, branching of the generalized dimensions arises. Moreover, the generalized dimensions have negative values in general. One can classify the random multifractal measures into three classes based on the discrepancy between the curves of generalized dimensions, which relate to $\mu([0,1]) = 1$ for conservative cascade and $\mu([0,1]) = \infty$ and $\mu([0,1]) = 0$ for non-conservative cascades. We also found that, one can perform the equivalent classification by investigating the location of the zero-point of $\tau(q)$ or the relative position either between the $f(\alpha)$ curve and the diagonal $f(\alpha) = \alpha$ or between the $f(q)$ curve and the $\alpha(q)$ curve. Therefore, we presented a phase diagram to characterize the classification procedure and distinguish the scaling properties between different classes. The occurrence of the branching phenomenon emerged in the curve of the generalized dimensions follows a two-step procedure. If the extreme value condition fits, the investigated measure belongs to Class I. Otherwise, if the generalized dimensions converge at point $q = 1$, the measure lies in Class II. The residual measures fall in Class III.

## Acknowledgements

This research was supported by the National Development Programming of Key and Fundamental Researches of China (No. G1999022103).

## References

[1]  P. Grassberger, Phys. Lett. A 97 (1983) 227.

[2]  H. G. E. Hentschel, I. Procaccia, Physica D 8 (1983) 435.

[3]  P. Grassberger, Phys. Lett. A 107, 101 (1985).

[4]  U. Frisch, G. Parisi, in: M. Gil, R. Benzi, G. Parisi, eds., *Turbulence and Predictability in Geophysical Fluid*

*Dynamics* (North-Holland, New York, 1985), p. 84.

[5] T. C. Halsey, M. H. Jensen, L. P. Kadanoff, I. Procaccia, B. I. Shraiman, Phys. Rev. A 33 (1986) 1141.

[6] B. B. Mandelbrot, Physica A 168 (1990) 95.

[7] B. B. Mandelbrot, C. J. G. Evertsz, Y. Hayakawa, Phys. Rev. A 42 (1990) 4528.

[8] R. H. Riedi, J. Math. Analysis Appl. 189 (1995) 462.

[9] R. H. Riedi, B. B. Mandelbrot, Adv. Appl. Math. 16 (1995) 132.

[10] B. B. Mandelbrot, C. J. G. Evertsz, in: A. Bunde, S.Halvin, eds., *Fractals and disordered systems* (Springer, Berlin, 1996), p. 366.

[11] B. B. Mandelbrot, J. Fluid Mech. 62 (1974) 331.

[12] W. X. Zhou, Y. J. Wang, Z. H. Yu, J. East China Uni. Sci. Tech. 26 (2000), to appear.

[13] W. X. Zhou, T. Wu, Z. H. Yu, J. East China Uni. Sci. Tech. 26 (2000), to appear.

[14] C. Amitrano, A. Coniglio, F. Di Liberto, Phys. Rev. Lett. 57 (1986) 1016.

[15] M. E. Cates, T. A. Witten, Phys. Rev. A 35 (1987) 1809.

[16] C. Meneveau, K. R. Sreenivasan, J. Fluid Mech. 224 (1991) 429.

[17] A. B. Chhabra, K. R. Sreenivasan, Phys. Rev. A 43 (1991) 1114.

[18] B. B. Mandelbrot, *The Fractal Geometry of Nature* (Freeman, New York, 1982).

[19] B. B. Mandelbrot, Pure Appl. Geophys. 131 (1989) 5.

[20] B. B. Mandelbrot, Physica A 163 (1990) 306.

[21] B. B. Mandelbrot, Proc. Roy. Soc. London A 434 (1991) 79.

[22] T. C. Halsey, M. Leibig, Phys. Rev. A 46 (1992) 7793.

[23] T. C. Halsey, in: L. Lam, ed., *Introduction to Nonlinear Physics* (Springer-Verlag, New York, 1997).

[24] B. B. Mandelbrot, in: L. Pietronero, ed., *Proceedings of the 1988 Erice Workshop on Fractals* (Plenum, London, 1990).

[25] C. Meneveau, K. R. Sreenivasan, Phys. Lett. A 137 (1989) 103.

[26] H. Chernoff, Ann. Math. Stat. 23 (1952) 495.

[27] J. D. Deuschel, D. W. Stroock, *Large Deviations* (Academic Press, New York, 1989).

[28] A. Dembo, O. Zeitouni, *Large Deviations: Techniques and Applications* (Springer, New York, 1998).

[29] Since the integration $\int_0^1 M^q x^M \, dM$ has a spot at the point $M = 0$ for negative $q$, this spot integration is the limit of the normal integration $\int_\varepsilon^1 M^q x^M \, dM$ when $\varepsilon \to 0^+$. To improve the accuracy of the evaluation, $\varepsilon$ should take a very small value, which leads to exponentially more work.

[30] W. X. Zhou, T. Wu, Z. H. Yu, *Features arising from randomly multiplicative measures*. (Submitted to Phys. Rev. E).

[31] W. X. Zhou, Z. H. Yu, J. Nonlinear Dynamics Sci. Tech. 7 (2000) 48.

[32] B. B. Mandelbrot, in: L. Pietronero, ed., *Fractals' Physical Origin and Properties* (Plenum, New York, 1989). p. 3

[33] B. B. Mandelbrot, Limit lognormal multifractal measures, in: E. Dotsman, Y Ne'eman, A. Voronel, eds., *Frontiers of Physics: Landau Memorial Conference* (Pergamon, New York, 1989). p. 309

[34] B. B. Mandelbrot, in: M. Rosenblatt, C. Van Atta, eds., *Statistical Models and Turbulence (Lecture Note in Physics, Vol. 12)* (Springer-verlag, New York, 1972)

[35] K. J. Falconer, T. C. O'Neil, Proc. R. Soc. Lond. A 452 (1996) 1433.

[36] S. D. Liu, S. S. Liu, *Introduction to Fractals and Fractal Dimensions* (meteorological Press, Beijing, 1993).

[37] W. X. Zhou, T. J. Zhao, T. Wu, Z. H. Yu, Chemical Engineering Journal 78 (2000) 193.

[38] W. X. Zhou, H. Liu, T. Wu, Z. H. Yu, Journal of Nonlinear Dynamics in Science and Technology 7 (2000) 90.